\documentclass[epj]{svjour}

\usepackage{epsfig}
\usepackage{geometry}
\usepackage{axodraw}
\usepackage{slashed}
\usepackage{amsmath}
\usepackage{amssymb}
\usepackage{amsfonts}

\usepackage{graphicx}
\usepackage{graphics}
\usepackage{fancyhdr}
\def\makeheadbox{{
 }}
\def\mytitle{My title} 
\def\myauthors{My name}  
\def\mytype{My type of session}
\def\mysession{My session}

\def\mytitle{R-Parity violating minimal supergravity at the LHC} 
\def\myauthors{Sebastian Grab}    
\def\mytype{Contributed Talk}    
\def\mysession{Colliders - SUSY Phenomenology}

\def\lsim{\raise0.3ex\hbox{$\;<$\kern-0.75em\raise-1.1ex\hbox{$\sim\;$}}}
\def\gsim{\raise0.3ex\hbox{$\;>$\kern-0.75em\raise-1.1ex\hbox{$\sim\;$}}}

\begin{document}
\title{R-Parity violating minimal supergravity at the LHC}
\author{B.~C.~Allanach\inst{1}
\thanks{\emph{Email:} b.c.allanach@damtp.cam.ac.uk},  %
  M.~A.~Bernhardt\inst{2}
\thanks{\emph{Email:} markus@th.physik.uni-bonn.de}, %
  H.~K.~Dreiner\inst{2}
\thanks{\emph{Email:} dreiner@th.physik.uni-bonn.de}, %
  S.~Grab\inst{2}
\thanks{Speaker, \emph{Email:} sgrab@th.physik.uni-bonn.de}, %
  C.~H.~Kom\inst{1}
\thanks{\emph{Email:} c.kom@damtp.cam.ac.uk}, %
\and
  P.~Richardson\inst{3}
\thanks{\emph{Email:} Peter.Richardson@durham.ac.uk}
}                     

\institute{DAMTP, University of Cambridge, Cambridge, UK
\and Physikalisches Institut, University of Bonn, Bonn, Germany
\and IPPP, University of Durham, Durham, UK}
%
\date{}
\abstract{
We consider the case where supersymmetry with broken R-parity is embedded in the
minimal supergravity model (mSUGRA). This alters the standard mSUGRA spectrum and
opens a wide range in parameter space, where the scalar tau is the lightest
supersymmetric particle, instead of the lightest neutralino. We study the resulting
LHC phenomenology. Promising signatures would be detached vertices from long-lived
staus, multi lepton final states and multi-tau final states. We investigate 
in detail the corresponding cross sections and decay rates in characteristic 
benchmark scenarios.
\PACS{
      {04.65.+e}{Supergravity}   \and
      {12.60.Jv}{Supersymmetric models } \and
      {14.80.Ly}{Supersymmetric partners of known particles}
     } 
} 
\maketitle
\section{R-Parity violating minimal supergravity model}
\label{intro}
The general renormalizable superpotential of the minimal supersymmetric extension 
of the SM includes lepton and baryon number violating terms \cite{Dreiner:1997uz}:
\begin{eqnarray}
W_{\not R_p} & = & \epsilon_{ab}\left[\frac{1}{2} \lambda_{ijk} L_i^aL_j^b
\bar E_k + \lambda'_{ijk} L_i^aQ_j^{bx}\bar D_{kx}\right] \nonumber \\
&&
+\epsilon_{ab}\kappa^i  L_i^aH_2^b
+\frac{1}{2}\epsilon_{xyz}\lambda''_{ijk}
\bar U_i^{\,x} \bar D_j^{\,y} \bar D_k^{\,z} \,.
\label{RPV_superpot}
\end{eqnarray}

\begin{figure}[t!]
\begin{center}
\unitlength=1in
\begin{picture}(3,2.3)
	  \put(-0.6,0){\epsfig{file=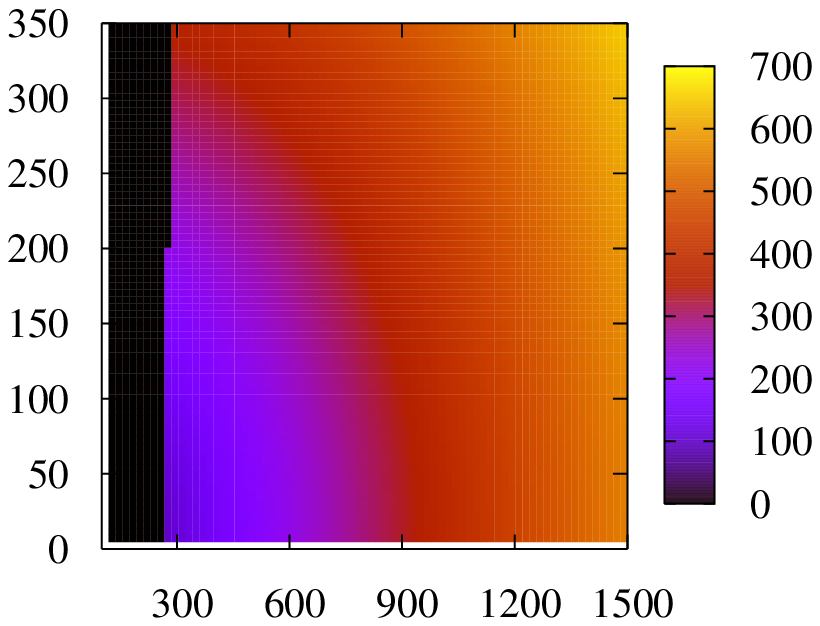, width=3.7in}}
	  \put(0.33,0.48){\epsfig{file=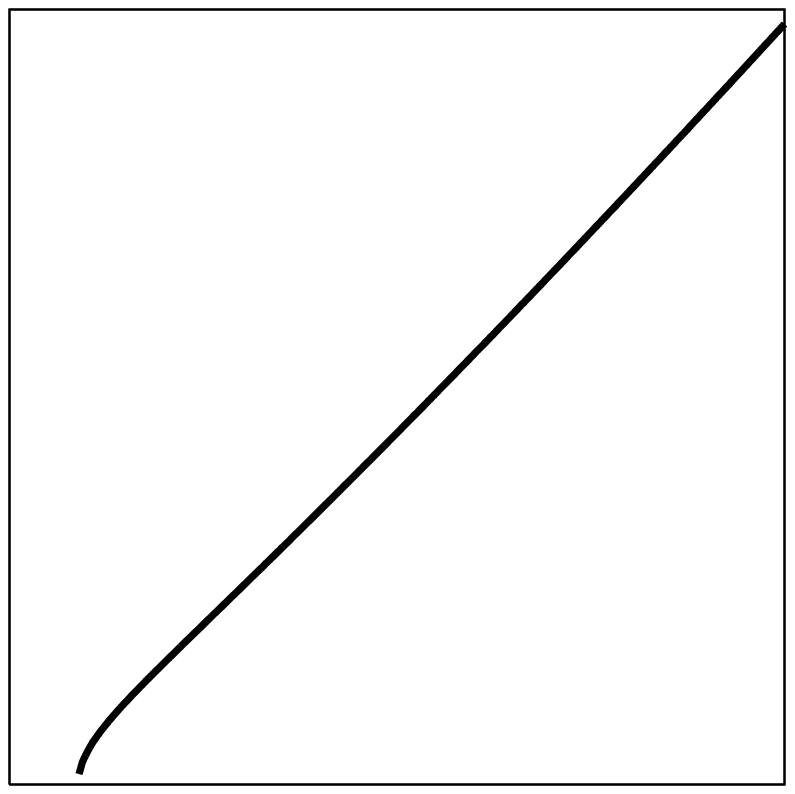, width=2.509in}}
	  \put(2.7,1.1){\rotatebox{90}{$m_{\tilde{\tau}}$ [GeV]}}
	  \put(1.3,0.2){\makebox(0,0){$M_{1/2}$ [GeV]}}
	  \put(0.0,1.2){\rotatebox{90}{$M_{0}$ [GeV]}}
	  \put(1.5,1.0){\makebox(0,0){\bf \White{$\boldsymbol{\tilde{\tau}}$\bf{--LSP}}}}
      \put(1.1,1.6){\makebox(0,0){\bf \White{$\boldsymbol{\tilde{\chi}^0_1}$\bf{--LSP}}}}
\end{picture}
\end{center}
\vspace{-0.6cm}
\caption{Parameter space around SPS1a, with $A_0 = -100$ GeV, $\tan \beta = 10$, and sgn($\mu$)=+1.
  The bar on the right displays the mass of the lightest stau $m_{\tilde{\tau}}$.   
  On the left side, we show in black the excluded  region due to tachyons and LEP2 Higgs exclusion bounds.
  The black contour distinguishes between areas with
  $\tilde{\tau}$-LSP and $\tilde{\chi}^0_1$-LSP \cite{Allanach:2006st}.}
\label{fig:triangle}
\end{figure}

The lepton and baryon number violating interactions in Eq. (\ref{RPV_superpot})
lead to rapid proton decay \cite{proton-decay}, if simultaneously present. 
Therefore the supersymmetric Lagrangean must respect an additional discrete symmetry.
The most widely studied scenario is R-Parity \cite{Farrar:1978xj}, $R_p$, for which $W_{\slashed{R}_p}=0$.
This scenario is conventionally named the minimal supersymmetric standard model (MSSM).
A detailed analysis of all discrete symmetries within the minimal supersymmetric extension 
of the SM can be found in Ref. \cite{Dreiner:2005rd}.  

The MSSM with $R_p$ has 124 free parameters \cite{Haber:1997if}, and without $R_p$ more than 200.
Thus it is mandatory to consider simpler models with well motivated boundary conditions
at the unification scale, $M_{GUT}$, because
such an extensive parameter space is intractable for a systematic phenomenological analyses at colliders. 
The most widely studied model is mSUGRA \cite{msugra} with radiative electroweak 
symmetry breaking \cite{Ibanez:1982fr} and conserved $R_p$. This model has only five parameters
(defined at $M_{GUT}$) 
\begin{equation}
M_0, \quad M_{1/2}, \quad A_0, \quad \tan\beta, \quad \text{sgn}(\mu).
\label{mSUGRA_param}
\end{equation} 
Here $M_0$ is the universal soft-breaking scalar mass, $M_{1/2}$ denotes the universal
gaugino mass, $A_0$ is the universal soft-breaking scalar interaction  
and $\tan\beta$ is the ratio of the vacuum expectation values of the two Higgs doublets.
The sign of the Higgs mixing parameter is sgn($\mu$). 

Fixing the five parameters (\ref{mSUGRA_param}) at the GUT scale allows us to compute the full supersymmetric spectrum 
and couplings at the electroweak scale through the renormalization group equations. A well studied
benchmark point is the SPS1a parameter set, $M_0 = 100$ GeV, $M_{1/2} = 250$ GeV, $A_0 = -100$ GeV,
$\tan \beta = 10$ and sgn($\mu$)=+1 \cite{Allanach:2002nj}. 

In Fig. \ref{fig:triangle}, we present the nature of the LSP 
around SPS1a for different values of $M_0$ and $M_{1/2}$ \cite{Allanach:2006st}.  
For cosmological reasons, a stable LSP has to be electrically and color neutral \cite{Ellis:1983ew}.
Therefore, 
the region with the lightest stau, $\tilde{\tau}$, is excluded if $R_p$ is conserved. If $R_p$ is violated,
the $\tilde{\tau}$ will decay into SM particles and does not contribute to dark matter anymore\footnote{
It was shown in \cite{Buchmuller:2007ui}, that small $\slashed{R}_p$ couplings can be consistent with 
primordial nucleosynthesis, thermal leptogenesis and gravitino dark matter.}.
Thus models with R-parity violation, $\slashed{R}_p$, reopen  
the $\tilde{\tau}$-LSP parameter space in Fig. \ref{fig:triangle}.

We take this as a motivation to add one $\slashed{R}_p$ coupling, 
\begin{equation}
\mathbf{\Lambda} \in \{\lambda_{ijk}, \lambda'_{ijk}, \lambda''_{ijk}\},
\end{equation}
to the five mSUGRA parameters (\ref{mSUGRA_param}). This leads to the $R_p$-violating mSUGRA model, which was 
considered in Ref. \cite{Allanach:2003eb}. In the following we investigate the phenomenology of 
$\tilde{\tau}$-LSP scenarios at the LHC.

\begin{figure}[h!]
  \begin{center}\scalebox{0.9}{
    \begin{picture}(210,70)(0,0)
      \ArrowLine(10,35)(60,35)
      \ArrowArc(80,35)(20,0,180)
      \ArrowArcn(80,35)(20,0,180)
      \ArrowLine(100,35)(150,35)
      \ArrowLine(200,65)(150,35)
      \ArrowLine(200,5)(150,35)
      \Vertex(60,35){1.5}
      \Vertex(100,35){1.5}
      \Vertex(150,35){1.5} 
      \put(35,40){$L_2$}
      \put(45,25){$\lambda'_{211}$}
      \put(80,60){$Q_1$}
      \put(80,5){$D^c_1$}
      \put(101,24){$(Y_D)^*_{11}$}
      \put(120,40){$H_D$}
      \put(160,32){$(Y_E)_{33}$}
      \put(195,55){$L_3$}
      \put(195,10){$E^c_3$}
    \end{picture}}
  \end{center}
  \vspace{-0.2cm}
  \caption{Dynamical generation of $\lambda_{233}$ via $\lambda'_{211}$.}
  \label{fig:DynGen_lambda}
\end{figure}
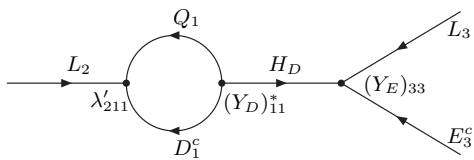

Note, that a $\slashed{R}_p$ coupling at the GUT scale, which violates one lepton number, will dynamically generate all
$\slashed{R}_p$ couplings at the EW scale, which violate the same lepton number. 
As an example we show the generation of $\lambda_{233}$ via $\lambda'_{211}$ 
in Fig. \ref{fig:DynGen_lambda}. 
The generated couplings may lead to 2-body decays, which can dominate the $\tilde{\tau}$ decays (see below). 
But this strongly depends on the scenario, i.e. $\mathbf{\Lambda}(M_{GUT})$.

\section{General hadron collider signatures}
\label{general_pheno}

\begin{table}
\caption{4-body decays of the $\tilde{\tau}$ via a non-vanishing $\lambda'_{211}$ coupling.}
\label{tab:4body_decays}      
\begin{tabular}{clclc}
\hline\noalign{\smallskip}
 & $\tilde{\tau}^- \rightarrow \tau^- \bar \nu_\mu \bar d d$ & & $\tilde{\tau}^- \rightarrow \tau^- \nu_\mu d \bar d$ & \\
 & $\tilde{\tau}^- \rightarrow \tau^- \mu^+ \bar u d$ & & $\tilde{\tau}^- \rightarrow \tau^- \mu^- u \bar d$ & \\
 & $\tilde{\tau}^- \rightarrow \nu_\tau \mu^- u \bar u$ & & $\tilde{\tau}^- \rightarrow \nu_\tau \nu_\mu d \bar u$ & \\
 & $\tilde{\tau}^- \rightarrow \nu_\tau \mu^- d \bar d$ & & $\tilde{\tau}^- \rightarrow \nu_\tau \bar \nu_\mu \bar u d$ & \\
\noalign{\smallskip}\hline
\end{tabular}
\end{table} 

For small $\slashed{R}_p$ couplings, $\mathbf{\Lambda} \lsim \mathcal{O}(10^{-3})$, 
sparticles like squarks will be mainly produced in pairs at the LHC. 
The sparticles will cascade down in 2-body decays to the $\tilde{\tau}$-LSP via the $R_p$ gauge interactions.
Since the $\tilde{\tau}$-LSP can decay only via a $\slashed{R}_p$ operator, the signatures at the LHC strongly depend on
$\mathbf{\Lambda}$. There are two distinct decay modes:
\begin{itemize}
\item {\bf 2-body decays}: If the $\tilde{\tau}$ directly couples to the $\slashed{R}_p$ operator, 
e.g. for $\lambda'_{311} \not = 0$,
then the $\tilde{\tau}$ will decay via 2-body decay into SM particles.
\item {\bf 4-body decays}: If the $\tilde{\tau}$ does not directly couple to the $\slashed{R}_p$ operator, 
e.g. for $\lambda'_{211} \not = 0$ (cf. Table \ref{tab:4body_decays}), then the $\tilde{\tau}$ 
will decay via a 4-body decay into SM particles. 
\end{itemize}

Promising signatures of $\slashed{R}_p$ mSUGRA models with a $\tilde{\tau}$-LSP are:
\begin{itemize}
\item {\bf detached vertices}: For $\mathbf{\Lambda} \lsim \mathcal{O}(10^{-6})$, one would observe a detached vertex,
if 2-body decays dominate. For 4-body decays the decay pattern is more involved, since 
the $\tilde{\tau}$ lifetime strongly depends on the specific scenario, due to the virtual sparticles.
For $\mathbf{\Lambda} \lsim \mathcal{O}(10^{-2}-10^{-4})$, 
then the observation of detached vertices is more likely.
\item {\bf multi lepton final states}: If the dominant $\tilde{\tau}$ decay is due to $\lambda_{ijk} \not = 0$,
then we will obtain one lepton and one neutrino (three leptons and one neutrino) 
for a $\tilde{\tau}$ 2-body decay (4-body decay). 
\item {\bf multi tau final states}: Supersymmetric decay chains can involve the lightest neutralino, which decays into the 
$\tilde{\tau}$-LSP and a $\tau$ via $\tilde{\chi}_1^0 \rightarrow \tilde{\tau}^+ \tau^- (\tilde{\tau}^- \tau^+)$.
4-body $\tilde{\tau}$ decays via a virtual neutralino ($\tilde{\tau}^- \rightarrow \tau^- \tilde{\chi}_1^{0*}$) 
lead also to an additional tau in the final state. Therefore, good tau identification is essential to identify
such scenarios.
\item{\bf like-sign dileptons}: Decays of two on-shell or virtual (Majorana) neutralinos into a lepton
and a slepton can lead to like-sign dilepton events. 
\end{itemize}

\begin{figure}[t!]
	  \scalebox{0.8}{
      \begin{picture}(0,0)
      \put(3,0){\epsfig{file=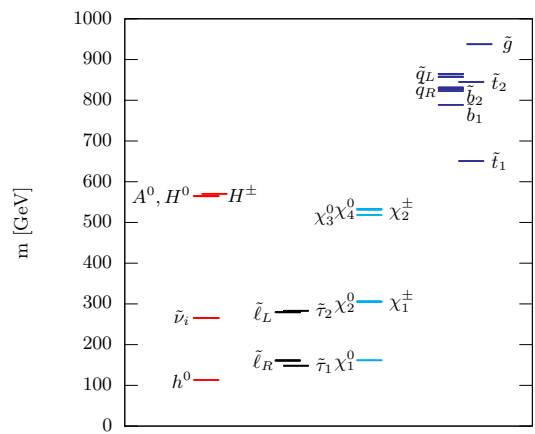}}
	  \end{picture}
		\setlength{\unitlength}{0.0200bp}%
		\begin{picture}(18000,10800)(0,0)%
		\put(2700,600){\makebox(0,0)[r]{\strut{} 0}}%
		\put(2700,1560){\makebox(0,0)[r]{\strut{} 100}}%
		\put(2700,2520){\makebox(0,0)[r]{\strut{} 200}}%
		\put(2700,3480){\makebox(0,0)[r]{\strut{} 300}}%
		\put(2700,4440){\makebox(0,0)[r]{\strut{} 400}}%
		\put(2700,5400){\makebox(0,0)[r]{\strut{} 500}}%
		\put(2700,6360){\makebox(0,0)[r]{\strut{} 600}}%
		\put(2700,7320){\makebox(0,0)[r]{\strut{} 700}}%
		\put(2700,8280){\makebox(0,0)[r]{\strut{} 800}}%
		\put(2700,9240){\makebox(0,0)[r]{\strut{} 900}}%
		\put(2700,10200){\makebox(0,0)[r]{\strut{} 1000}}%
		\put(600,5400){\rotatebox{90}{\makebox(0,0){\strut{}m [GeV]}}}%
		\put(4344,1688){\makebox(0,0){$h^0$}}%
		\put(3856,6010){\makebox(0,0){$A^0,H^0$}}%
		\put(5784,6085){\makebox(0,0){$H^{\pm}$}}%
		\put(4344,3151){\makebox(0,0){$ \tilde \nu_i$}}%
		\put(6264,3284){\makebox(0,0){$ \tilde\ell_L$}}%
		\put(6264,2149){\makebox(0,0){$ \tilde\ell_R$}}%
		\put(7704,3318){\makebox(0,0){$ \tilde\tau_2$}}%
		\put(7704,2023){\makebox(0,0){$ \tilde\tau_1$}}%
		\put(8184,2151){\makebox(0,0){$ \chi_1^0$}}%
		\put(8184,3521){\makebox(0,0){$ \chi_2^0$}}%
		\put(7704,5577){\makebox(0,0){$ \chi_3^0$}}%
		\put(8184,5719){\makebox(0,0){$ \chi_4^0$}}%
		\put(9528,3542){\makebox(0,0){$ \chi_1^{\pm}$}}%
		\put(9528,5702){\makebox(0,0){$ \chi_2^{\pm}$}}%
		\put(11832,8712){\makebox(0,0){$ \tilde t_2$}}%
		\put(11832,6848){\makebox(0,0){$ \tilde t_1$}}%
		\put(10104,8902){\makebox(0,0){$ \tilde q_L$}}%
		\put(10104,8517){\makebox(0,0){$ \tilde q_R$}}%
		\put(11256,8406){\makebox(0,0){$ \tilde b_2$}}%
		\put(11256,7977){\makebox(0,0){$ \tilde b_1$}}%
		\put(12024,9601){\makebox(0,0){$ \tilde g$}}%
		\end{picture}}
	  \vspace{-0.3cm}
	\caption{Sparticle mass spectrum for benchmark scenario BC1 (BC2), i.e.
			 $A_0=M_0=0$ GeV, $M_{1/2}=400$ GeV, $\tan \beta=13$, 
			 sgn($\mu$)=+1, $\lambda_{121}(M_{GUT})=0.032$ ($\lambda'_{311}(M_{GUT})=3.5\times10^{-7}$) \cite{Allanach:2006st}.
			 \label{fig:benchI}} 
\end{figure}
 
For $\mathbf{\Lambda} \gsim \mathcal{O}(10^{-2})$, the $\slashed{R}_p$ operator can significantly change the decay chains 
of the sparticles. Also the mass spectrum can be significantly affected \cite{Allanach:2006st}. 
Furthermore single sparticle production may dominate, e.g., single slepton production 
for $\lambda'_{ijk}\not=0$ \cite{ResSlep}.

\section{Stau LSP phenomenology at the LHC.}
\label{pheno_at_LHC} 

\subsection{Benchmark scenarios}
\label{benchmarks}

In Ref. \cite{Allanach:2006st}, four benchmark scenarios, BC1--BC4, within the framework of $\slashed{R}_p$ mSUGRA 
were proposed. Three scenarios, BC1, BC2, BC4, contain a $\tilde{\tau}$-LSP and BC3 a sneutrino-LSP.
In the following we will consider two of these scenarios, namely:
\begin{itemize}
\item {\bf BC1}: $M_0=A_0=0$ GeV, $M_{1/2}=400$ GeV, $\tan \beta=13$, 
sgn($\mu$)=+1, $\lambda_{121}=0.032$ at $M_{GUT}$.
\item {\bf BC2}: $M_0=A_0=0$ GeV, $M_{1/2}=400$ GeV, $\tan \beta=13$, 
sgn($\mu$)=+1, $\lambda'_{311}=3.5 \times 10^{-7}$ at $M_{GUT}$.
\end{itemize} 
Both scenarios differ only by the non-vanishing $\slashed{R}_p$ operator. The mass spectrum is shown in 
Fig. \ref{fig:benchI}. The $\tilde{\tau}$ is the LSP with $m_{\tilde{\tau}}=148$ GeV.
The spectrum is calculated with an $\slashed{R}_p$ version of \texttt{SOFTSUSY} \cite{RPV_softsusy}.
The effect of the $\slashed{R}_p$ operators on the sparticle masses is smaller than 1 GeV and thus negligible. 
Note that for BC1 and BC2, the NLSP to NNNLSP are nearly degenerate in mass 
($m_{\tilde{e}_R}=161$ GeV, $m_{\tilde{\mu}_R}=161$ GeV, $m_{\tilde{\chi}^0_1}=162$ GeV).  

\subsection{Signal rates at the LHC}
\label{signal_rates}

We now investigate the signatures at the LHC at parton level. We use a modified version of \texttt{HERWIG} \cite{herwig}, 
which contains all 2- and 4-body decays of the $\tilde{\tau}$-LSP. 
All other decay rates of sparticles are calculated with \\ \texttt{ISAWIG1.200} and \texttt{ISAJET7.64} \cite{Paige:2003mg}.

\begin{table}
\caption{Signal rates for benchmark scenario BC1. The numbers of electrons or muons of positive
charge (first row) and negative charge (second row) are shown. We also give the numbers of taus
with positive (third row) and negative charge (fourth row). Furthermore, we mention, if there 
is missing $p_T$ (neglecting that from $\tau$ decay) due to neutrinos 
(fifth row) and we present the probability of the different signatures
(sixth row). Each event is accompanied by 2--4 jets. $\sigma_{tot}=4.8\times10^{3}$ fb.}
\label{tab:signal_BC1}      
\begin{tabular}{cccccc}
\hline\noalign{\smallskip}
$e^+$ or $\mu^+$ & $e^-$ or $ \mu^-$ & $\tau^+$ & $\tau^-$ & $\slashed{p}_T$ & event fraction  \\
\noalign{\smallskip}\hline\noalign{\smallskip}
2 & 2 & 2 & 2 & yes & $35$ \% \\
	3 & 2 & 2 & 2 & yes & $12$ \% \\
	2 & 3 & 2 & 2 & yes & $8.3$ \% \\
	3 & 3 & 2 & 2 & yes & $7.3$ \% \\
	2 & 2 & 2 & 1 & yes & $4.7$ \% \\
	2 & 2 & 3 & 2 & yes & $4.3$ \% \\
	2 & 2 & 3 & 3 & yes & $1.4$ \% \\
	4 & 3 & 2 & 2 & yes & $1.1$ \% \\
\noalign{\smallskip}\hline
\end{tabular}
\end{table}

\begin{table}
\caption{Dominant branching ratios (BR) of the LSP, $\cdots$, NNNLSP for benchmark scenario BC1.
$\slashed{R}_p$ BRs are bold face \cite{Allanach:2006st}.}
\label{tab:BR_BC1}      
\begin{tabular}{cclclc}
\hline\noalign{\smallskip}
 & mass [GeV] & channel & BR & channel & BR  \\
\noalign{\smallskip}\hline\noalign{\smallskip}
	$\tilde{\tau}_1^-$& 148 & $\mu^+ \bar \nu_e e^- \tau^-$ & {\bf 32} \%  
	& $e^+ \bar \nu_\mu e^- \tau^-$ & {\bf 32} \% \\
	& &$\mu^-\nu_e e^+ \tau^-$ & {\bf 18} \%
	& $e^-\nu_\mu e^+\tau^-$ & {\bf 18} \% \\ 
	$\tilde{e}_R^-$ & 161 & $e^-\nu_\mu$ & {\bf 50} \% 
	&$\mu^-\nu_e$ & {\bf 50} \% \\ 
	$\tilde{\mu}_R^-$& 161 & $\tilde{\tau}^+ \mu^- \tau^-$ & 51 \% 
	& $\tilde{\tau}^- \mu^- \tau^+$ & 49 \% \\
	$\tilde{\chi}_1^0$ & 162 & $\tilde{\tau}_1^+ \tau^-$ & 50 \% 
	& $\tilde{\tau}_1^- \tau^+$ & 50 \% \\
\noalign{\smallskip}\hline
\end{tabular}
\end{table}

\begin{figure}[t!]
\begin{center}
\unitlength=1in
\scalebox{0.9}{
\begin{picture}(3,2.3)
	  \put(-0.05,0.3){\epsfig{file=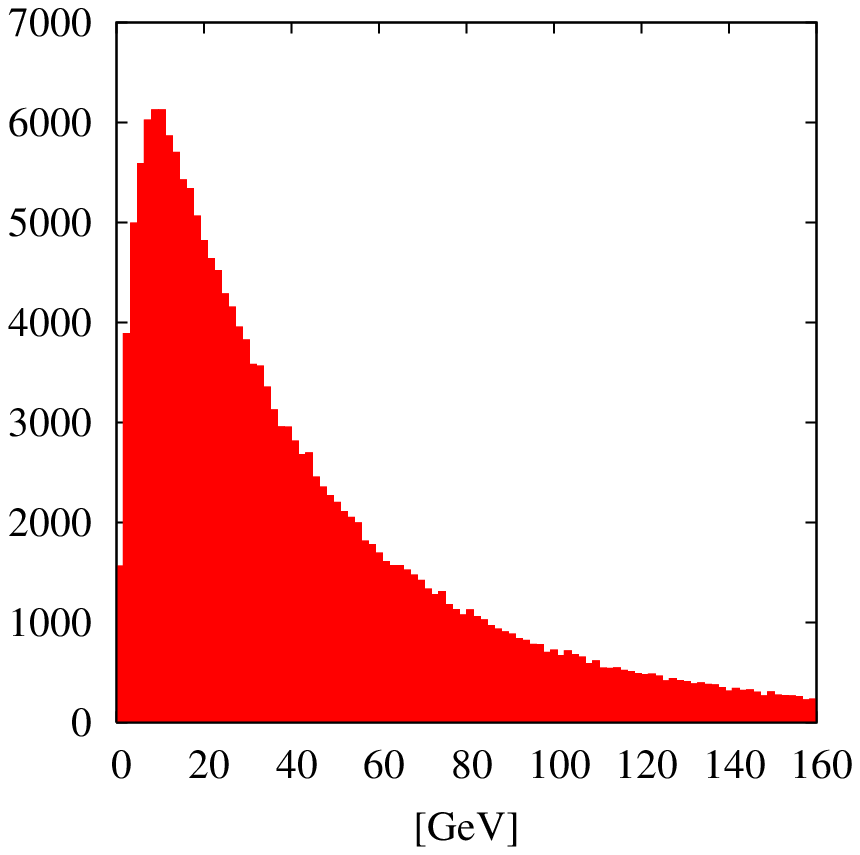,bb = 105 70 325 290, clip=true, width=1.52in}}
	  \put(1.6,0.3){\epsfig{file=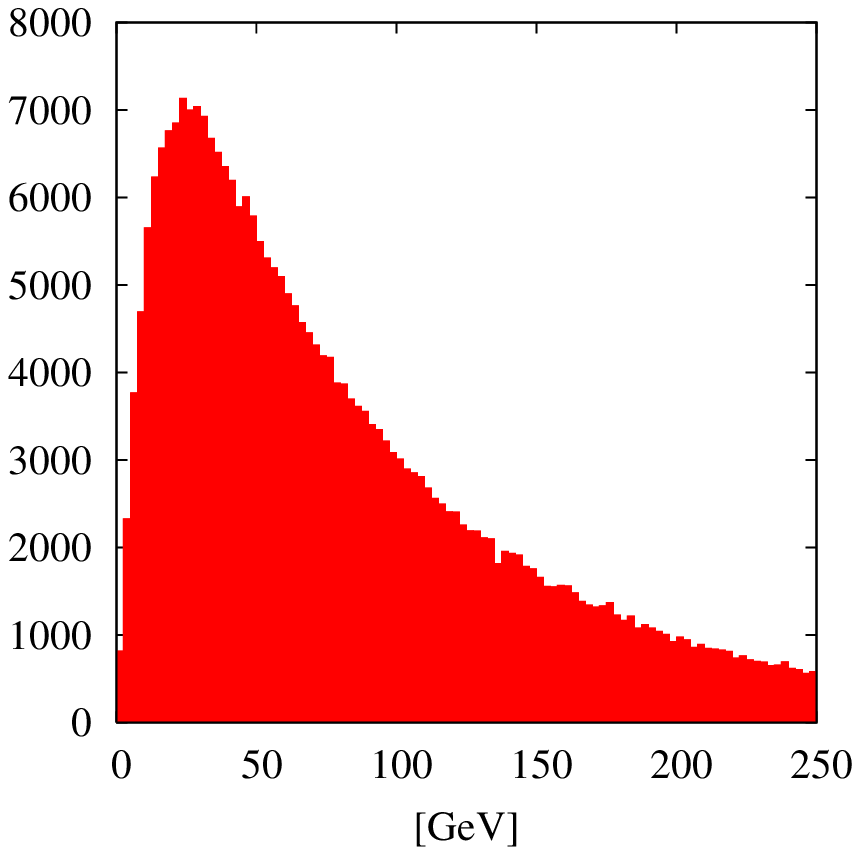,bb = 100 70 320 290, clip=true, width=1.53in}}
	  \put(-0.17,0.6){\rotatebox{90}{number of events}}
	  \put(1.5,0.6){\rotatebox{90}{number of events}}
	  \put(0.4,0.1){{$p_T$ [GeV]}}
	  \put(2.1,0.1){{$p_T$ [GeV]}}
	  \put(0.55,1.6){{$\tau$ $p_T$ in BC1}}
	  \put(2.3,1.6){{$\nu$ $p_T$ in BC1}}
\end{picture}}
\end{center}
\vspace{-0.5cm}
\caption{$p_T$ distribution of the $\tau$ from $\tilde{\tau} \rightarrow \tau + X$ (left) and
		 $p_T$ distribution of the neutrinos (right) in benchmark scenario BC1.}
\label{fig:pT_BC1}
\end{figure}

We show the signal rates of the benchmark scenario BC1 in Table \ref{tab:signal_BC1}. 
The total cross section for all sparticle pair production processes is $\sigma_{tot}=4.8\times10^{3}$ fb.
Each event is accompanied by four or more electrons or muons in the final state.  
These leptons allow for triggering and for reconstruction of the event. Most of these leptons
originate from the 4-body decays of the $\tilde{\tau}$-LSP via $\lambda_{121}$, see Table \ref{tab:BR_BC1}.  
We also observe in Table \ref{tab:BR_BC1}, that the $\tilde{e}_R$ decays into a lepton and neutrino, since it
directly couples to the $L_1L_2\bar E_1$ operator. Each 4-body decay and each $\tilde{\chi}_1^0$ decay is 
accompanied by a tau, which leads to 4 taus in most of the events. Thus a good tau identification would
help to identify this scenario. The low-$p_T$ taus will be invisible, yet the high energy tail, $p_T\geq 30$ GeV,
see Fig. \ref{fig:pT_BC1}, is useful. The small number of jets (2-4 jets) give only a small combinatorial
background for tau identification via hadronic decays. Finally, each 4-body decay includes a final-state neutrino, 
resulting in missing transverse momentum, $\slashed{p}_T$, shown in Fig. \ref{fig:pT_BC1}. This is however reduced
compared to the $R_p$-MSSM, where the $\slashed{p}_T$ peaks roughly around 100 GeV.

\begin{table}
\caption{Same as Table \ref{tab:signal_BC1}, but for benchmark scenario BC2.
Each event is accompanied by 6-8 jets.}
\label{tab:signal_BC2}      
\begin{tabular}{cccccc}
\hline\noalign{\smallskip}
$e^+$ or $\mu^+$ & $e^-$ or $ \mu^-$ & $\tau^+$ & $\tau^-$ & $\slashed{p}_T$ & event fraction  \\
\noalign{\smallskip}\hline\noalign{\smallskip}
	 0 & 0 & 1 & 1 & no & 14 \% \\
	 0 & 0 & 2 & 0 & no & 7.1 \% \\
	 0 & 0 & 0 & 2 & no & 6.8 \% \\
	 1 & 0 & 1 & 1 & yes & 6.5 \% \\
	 0 & 0 & 1 & 1 & yes & 4.5 \% \\
	 1 & 0 & 0 & 2 & yes & 3.3 \% \\
	 1 & 0 & 2 & 0 & yes & 3.2 \% \\
	 1 & 1 & 1 & 1 & yes & 2.4 \% \\
\noalign{\smallskip}\hline
\end{tabular}
\end{table}

\begin{table}
\caption{Same as Tab. \ref{tab:BR_BC1}, but for benchmark scenario BC2 \cite{Allanach:2006st}.}
\label{tab:BR_BC2}      
\begin{tabular}{cclclc}
\hline\noalign{\smallskip}
 & mass [GeV] & channel & BR & channel & BR  \\
\noalign{\smallskip}\hline\noalign{\smallskip}
	$\tilde{\tau}_1^-$ & 148 & $\bar{u} d$ & {\bf 100} \% & & \\
	$\tilde{e}_R^-$ & 161 & $\tilde{\tau}_1^+ e^-\tau^-$ & 51 \% & 
	$\tilde{\tau}_1^- e^- \tau^+$ & 49 \% \\ 
	$\tilde{\mu}_R^-$ & 161 & $\tilde{\tau}_1^+ \mu^- \tau^-$ & 51 \% & 
	$\tilde{\tau}_1^- \mu^- \tau^+$ & 49 \% \\ 
	$\tilde{\chi}_1^0$ & 162 & $\tilde{\tau}_1^+ \tau^-$ & 50 \%
	&$\tilde{\tau}_1^- \tau^+$ & 50 \% \\
\noalign{\smallskip}\hline
\end{tabular}
\end{table}

\begin{figure}[t!]
\begin{center}
\unitlength=1in
\scalebox{0.9}{
\begin{picture}(3,2.3)
	  \put(-0.05,0.3){\epsfig{file=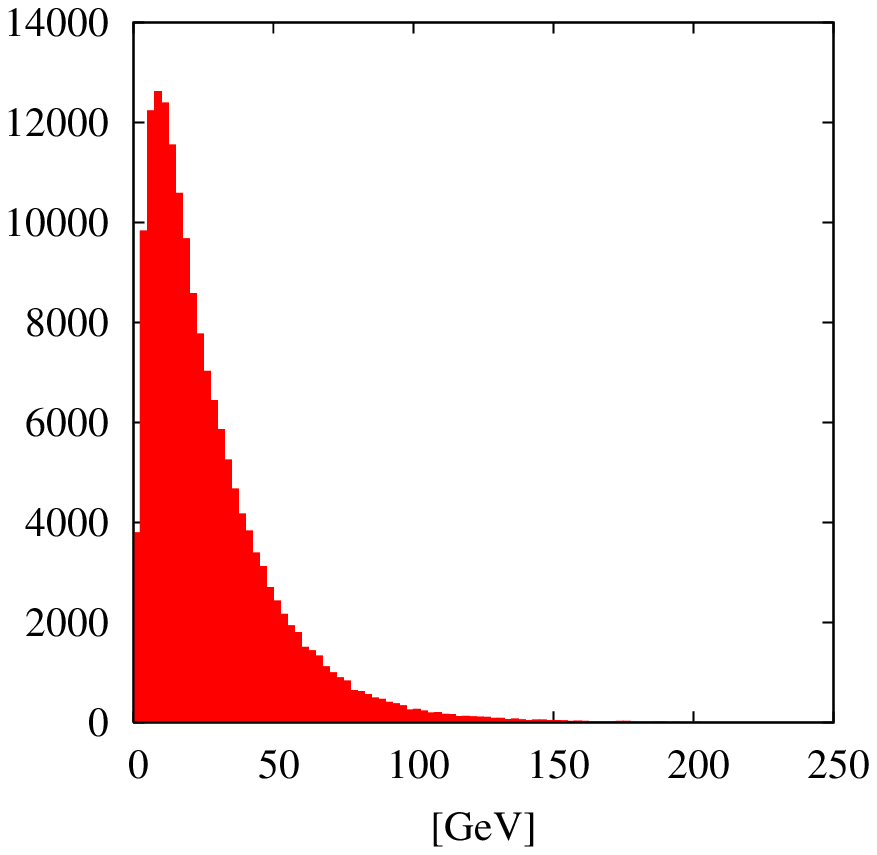,bb = 110 70 325 290, clip=true, width=1.5in}}
	  \put(1.6,0.3){\epsfig{file=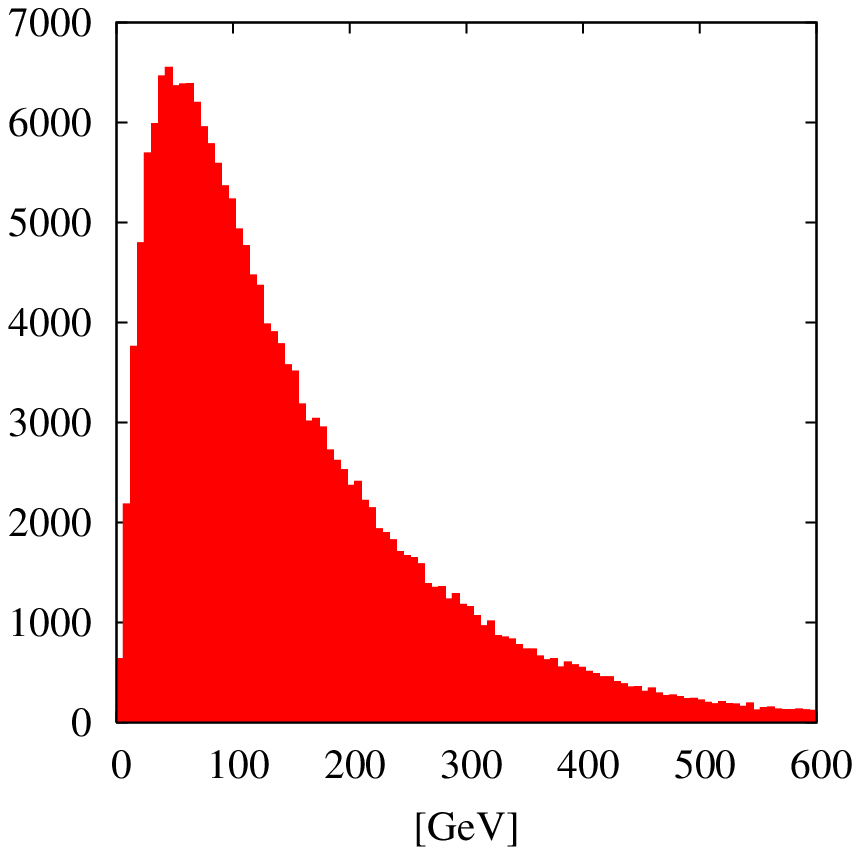,bb = 100 70 320 290, clip=true, width=1.54in}}
	  \put(-0.17,0.6){\rotatebox{90}{number of events}}
	  \put(1.5,0.6){\rotatebox{90}{number of events}}
	  \put(0.4,0.1){{$p_T$ [GeV]}}
	  \put(2.1,0.1){{$p_T$ [GeV]}}
	  \put(0.55,1.6){{$\tau$ $p_T$ in BC2}}
	  \put(2.1,1.6){{d-jet $p_T$ in BC2}}
\end{picture}}
\end{center}
\vspace{-0.5cm}
\caption{$p_T$ distribution of the $\tau$ from $\tilde{\chi}_1^0\rightarrow \tau \tilde{\tau}$ (left) and
		 $p_T$ distribution of the d-jets from $\tilde{\tau} \rightarrow u d$ (right) in benchmark scenario BC2.}
\label{fig:pT_BC2}
\end{figure}

We now move on to BC2, where we have $\lambda'_{311} = 3.5 \times 10^{-7}$ at $M_{GUT}$.
This changes the LHC signatures in a significant way, 
as can be seen in Table \ref{tab:signal_BC2}. Most of the electrons and muons are absent. 
The $\tilde{\tau}$-LSP is completely dominated by 2-body decays into two jets, see Table \ref{tab:BR_BC2}.
Unlike in BC1, the LSP decays are neutrinoless, meaning that SUSY events do not necessarily have the
classical signature of missing transverse momentum. Most events are accompanied by two
taus, which originate mainly from $\tilde{\chi}_1^0$ decays, see Table \ref{tab:BR_BC2}. These taus
are much harder to identify compared to BC1. On the one hand, the average $p_T$ values of the taus are smaller, as
can be seen in Fig. \ref{fig:pT_BC2}. On the other hand, the additional 6-8 jets in each event makes 
their identification difficult.
But there will be a detached vertex, since the decay length in the 
rest frame of the $\tilde{\tau}$-LSP is $c\tau_{\tilde{\tau}} \approx 0.3$ mm. The two jets 
from the 2-body decay make it possible to reconstruct the $\tilde{\tau}$-LSP mass. 
We show the $p_T$ distribution of one of these jets in Fig. \ref{fig:pT_BC2}.  

\section{Summary and conclusion}
\label{summary}
We have shown that minimal supergravity (mSUGRA) models often provide a 
scalar tau instead of the lightest neutralino as the lightest supersymmetric particle (LSP).
If R-Parity is violated, the stau will decay into SM particles and does not contribute to dark matter. 
This reopens a large part of the mSUGRA parameter space, where the stau is the LSP.
We have analysed in detail the LHC phenomenology of two benchmark scenarios with a stau-LSP. 
Although both scenarios posses the same mass spectrum, their signatures are different. 
In the first benchmark scenario, BC1, the stau-LSP decays via a 4-body decay into a tau,
two leptons and one neutrino.
In the second benchmark scenario, BC2, the stau-LSP decays into two jets. 
We have to be prepared to observe SUSY events at the LHC,
which are different from the standard signatures. In particular SUSY events 
must not have large missing transverse momentum. 

We conclude that further investigations of stau-LSP models from the phenomenological as well as
from the experimental side are required. We have to classify and analyse all possibilities
of the R-Parity violating mSUGRA parameter space, if we want to discover supersymmetry 
and its connection to physics at the GUT scale.

\section{Acknowledgments}
\label{acknowledgments}
The authors want to thank Klaus Desch and Sebastian Fleischmann for helpful discussions.
MB thanks DAMTP, Cambridge, and IPPP, Durham, for repeatedly offered, warm hospitality.
SG thanks the Deutsche Telekom Stiftung for financial support.
This work has been partially supported by STFC.


\begin{thebibliography}{999}

\bibitem{Dreiner:1997uz}
  H.~K.~Dreiner,
  arXiv:hep-ph/9707435.

\bibitem{proton-decay}
J.~L.~Goity and M.~Sher,
  Phys.\ Lett.\ B {\bf 346} (1995) 69
  [Erratum-ibid.\ B {\bf 385} (1996) 500];
A.~Y.~Smirnov and F.~Vissani,
  Phys.\ Lett.\ B {\bf 380} (1996) 317;
A.~Y.~Smirnov and F.~Vissani,
Nucl.\ Phys.\ B {\bf 460} (1996) 37.
      
\bibitem{Farrar:1978xj}
  G.~Farrar and P.~Fayet,
  Phys.\ Lett.\ B {\bf 76} (1978) 575.

\bibitem{Dreiner:2005rd}
  L.~E.~Ibanez and G.~G.~Ross,
  Nucl.\ Phys.\  B {\bf 368} (1992) 3;
  H.~K.~Dreiner, C.~Luhn and M.~Thormeier,
  Phys.\ Rev.\  D {\bf 73}, (2006) 075007.
  
\bibitem{Haber:1997if}
  H.~E.~Haber,
  Nucl.\ Phys.\ Proc.\ Suppl.\  {\bf 62} (1998) 469.
  
 \bibitem{msugra}
A.~Chamseddine, R.~Arnowitt and P.~Nath,
  Phys.\ Rev.\ Lett.\  {\bf 49} (1982) 970;
L.~Alvarez-Gaume, M.~Claudson and M.~Wise,
  Nucl.\ Phys.\ B {\bf 207} (1982) 96;
L.~Ibanez,
  Phys.\ Lett.\ B {\bf 118} (1982) 73.
S.~Soni and H.~Weldon,
  Phys.\ Lett.\ B {\bf 126} (1983) 215.
L.~Hall, J.~Lykken and S.~Weinberg,
  Phys.\ Rev.\ D {\bf 27} (1983) 2359.

\bibitem{Ibanez:1982fr}
  L.~Ibanez and G.~Ross,
  Phys.\ Lett.\ B {\bf 110} (1982) 215.

\bibitem{Allanach:2002nj}
  B.~C.~Allanach {\it et al.},
  arXiv:hep-ph/0202233.

\bibitem{Allanach:2006st}
  B.~C.~Allanach, M.~A.~Bernhardt, H.~K.~Dreiner, C.~H.~Kom and P.~Richardson,
  Phys.\ Rev.\  D {\bf 75} (2007) 035002.
  
\bibitem{Ellis:1983ew}
  J.~Ellis, \textit{et al.},
  Nucl.\ Phys.\ B {\bf 238} (1984) 453.
  
\bibitem{Buchmuller:2007ui}
  W.~Buchmuller, L.~Covi, K.~Hamaguchi, A.~Ibarra and T.~Yanagida,
  JHEP {\bf 0703} (2007) 037.

\bibitem{Allanach:2003eb}
  B.~C.~Allanach, A.~Dedes and H.~K.~Dreiner,
  Phys.\ Rev.\  D {\bf 69} (2004) 115002
  [Erratum-ibid.\  D {\bf 72} (2005) 079902].
  
\bibitem{ResSlep}
  S.~Dimopoulos and L.~J.~Hall, Phys.\ Lett.\  B {\bf 207} (1988) 210;  
  S.~Dimopoulos, R.~Esmailzadeh, L.~J.~Hall and G.~D.~Starkman,
  Phys.\ Rev.\  D {\bf 41} (1990) 2099;
  H.~K.~Dreiner, S.~Grab, M.~Kramer and M.~K.~Trenkel,
  Phys.\ Rev.\  D {\bf 75} (2007) 035003.
  
\bibitem{Paige:2003mg}
  F.~E.~Paige, S.~D.~Protopopescu, H.~Baer and X.~Tata,
  arXiv:hep-ph/0312045.
  
\bibitem{herwig}
G.~Corcella {\em et al}, JHEP {\bf 0101} (2001) 010;
S. Moretti, K. Odagiri, P. Richardson, M.H. Seymour and B.R. Webber,
 JHEP {\bf 0204} (2002) 028.
  
\bibitem{RPV_softsusy}
B.~C.~Allanach, M.~A.~Bernhardt,
preprint: BONN-TH-2007-06.

\end{thebibliography}
\end{document}